\documentclass[conference]{IEEEtran}

\usepackage{flushend}

\usepackage{cite}
\usepackage{amsmath,amssymb,amsfonts}
\usepackage{algorithmic}
\usepackage{graphicx}
\usepackage{textcomp}
\usepackage{xcolor}

\def\BibTeX{{\rm B\kern-.05em{\sc i\kern-.025em b}\kern-.08em
    T\kern-.1667em\lower.7ex\hbox{E}\kern-.125emX}}
    
\begin{document}

\title{Axial-LOB: High-Frequency Trading with \\Axial Attention}

\author{
\IEEEauthorblockN{Damian Kisiel}
\IEEEauthorblockA{\textit{Department of Computer Science} \\
\textit{University College London}\\
London, United Kingdom \\
d.kisiel@cs.ucl.ac.uk}
\and
\IEEEauthorblockN{Denise Gorse}
\IEEEauthorblockA{\textit{Department of Computer Science} \\
\textit{University College London}\\
London, United Kingdom \\
d.gorse@cs.ucl.ac.uk}
}

\maketitle

\begin{abstract}
Previous attempts to predict stock price from limit order book (LOB) data are mostly based on deep convolutional neural networks. Although convolutions offer efficiency by restricting their operations to local interactions, it is at the cost of potentially missing out on the detection of long-range dependencies. Recent studies address this problem by employing additional recurrent or attention layers that increase computational complexity. In this work, we propose \emph{Axial-LOB}, a novel fully-attentional deep learning architecture for predicting price movements of stocks from LOB data. By utilizing gated position-sensitive axial attention layers our architecture is able to construct feature maps that incorporate global interactions, while significantly reducing the size of the parameter space. Unlike previous works, Axial-LOB does not rely on hand-crafted convolutional kernels and hence has stable performance under input permutations and the capacity to incorporate additional LOB features. The effectiveness of Axial-LOB is demonstrated on a large benchmark dataset, containing time series representations of millions of high-frequency trading events, where our model establishes a new state of the art, achieving an excellent directional classification performance at all tested prediction horizons.
\end{abstract}


\section{Introduction}

Recent advances in processing power and wider access to market data have allowed machine learning techniques \cite{b19}, including deep learning (DL) architectures \cite{b20}, to be effectively applied to financial data prediction. In particular, research in high-frequency trading (HFT) has benefited greatly from the wider availability of highly granular microstructure data \cite{b22, b23}. Despite these large volumes of data, however, the main challenges remain, namely that financial time series are characterized by complex dynamics, non-stationarity, and very low signal to noise ratios, making them notoriously difficult to forecast. Traditional methods such as ARIMA \cite{b1} and VAR \cite{b2} attempt to cope with this problem by relying on carefully hand-engineered features. Modern data-driven techniques, on the other hand, take a more agnostic approach, with deep learning models being able to learn the underlying mechanisms driving high-frequency trends and uncover predictive features directly from data.

Convolutional neural networks (CNNs) have been the most heavily exploited DL architectures in this context, due to their generalization ability and high efficiency, achieved through parameter sharing. For example, the works of \cite{b3, b21} demonstrate their effectiveness to predict stock price movements from LOB data using only a few convolutional layers. By design, however, CNNs can only aggregate information within a local region, which prevents them from modelling long-range dependencies. Several architectures have been recently developed to address this limitation in the context of high-frequency trading. DeepLOB \cite{b4} combines convolutional layers with a long short-term memory network (LSTM), with an additional attention mechanism in \cite{b5}, to model long-range interactions. Another line of work uses atrous (dilated) convolutions \cite{b7} to extend the receptive field by inserting holes between the kernel elements. However, these techniques introduce additional problems: stacking CNNs with LSTMs increases complexity and processing times, while dilated kernels throw away information by skipping over some input elements. It should also be noted that previous works extract useful feature maps from limit order books (LOBs) by carefully tuning all parts of the convolutional network, including the shape of kernels, strides, and the number of filters at each layer, which introduces extra hyperparameters and makes the model highly dependent on the ordering of the input features.

\begin{figure*}[t]
\centerline{\includegraphics[scale=0.53]{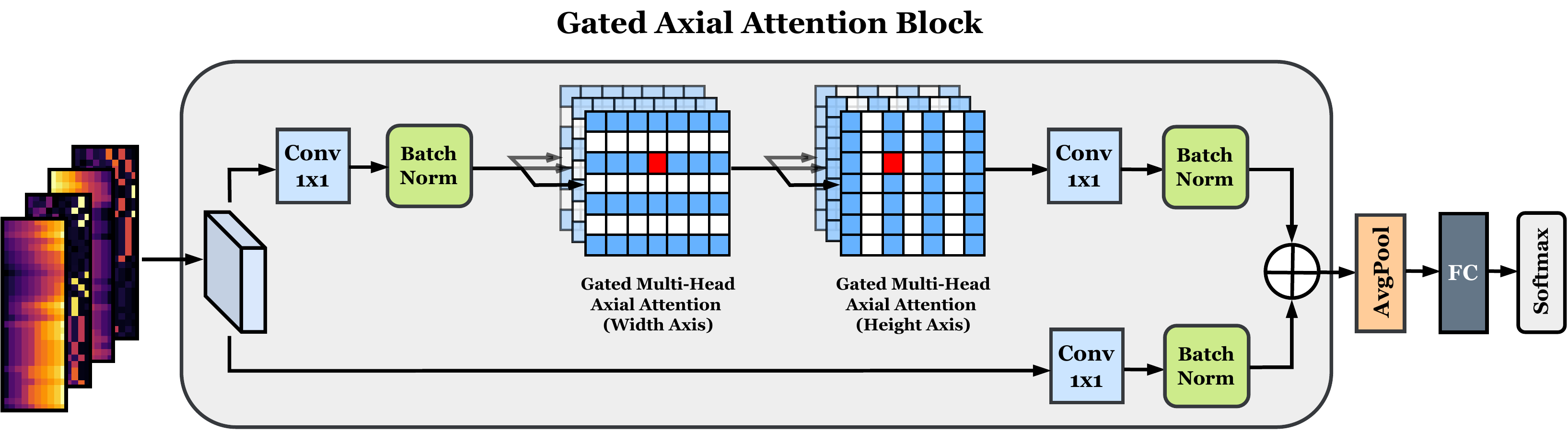}}
\caption{The Axial-LOB model architecture. Heatmaps show changes in prices and volumes on the bid and ask sides of the limit order book. The gated axial attention block is the main component of the architecture, implementing two consecutive gated axial attention operations (along width and height axes). For a more detailed look at the gated attention mechanism, refer to Fig.~\ref{fig3}. The 1x1 convolutions are used to manage model complexity via channel-wise pooling.}
\label{fig2}
\end{figure*}

In this work, we use a recently-proposed attentional architecture \cite{b13}, that does not require such hand-crafted spatial convolutions, as the basis for \emph{Axial-LOB}, a new model for the prediction of stock price movements from LOB data. In order to learn long-range dependencies, Axial-LOB uses axial attention layers that factorize the standard 2D attention mechanism into two 1D self-attention blocks (Fig.~\ref{fig2}), allowing the recovery of the global receptive field in a computationally efficient manner. Additionally, gated positional embeddings are used within the attention mechanisms, which enable the model to utilize and control the amount of position-dependent interactions. The effectiveness of Axial-LOB is demonstrated on the publicly available benchmark LOB dataset known as \emph{FI-2010}. Experimental results show that Axial-LOB achieves excellent directional classification performance in relation to mid-price movements. It sets a new state of the art for all prediction horizons, improving over the best prior work. Moreover, it has a much lower model complexity and demonstrates stable performance under permutations of the input data.

\section{Background \& Related Work}

\subsection{Limit Order Book}

\begin{figure}[b]
\centerline{\includegraphics[scale=0.63]{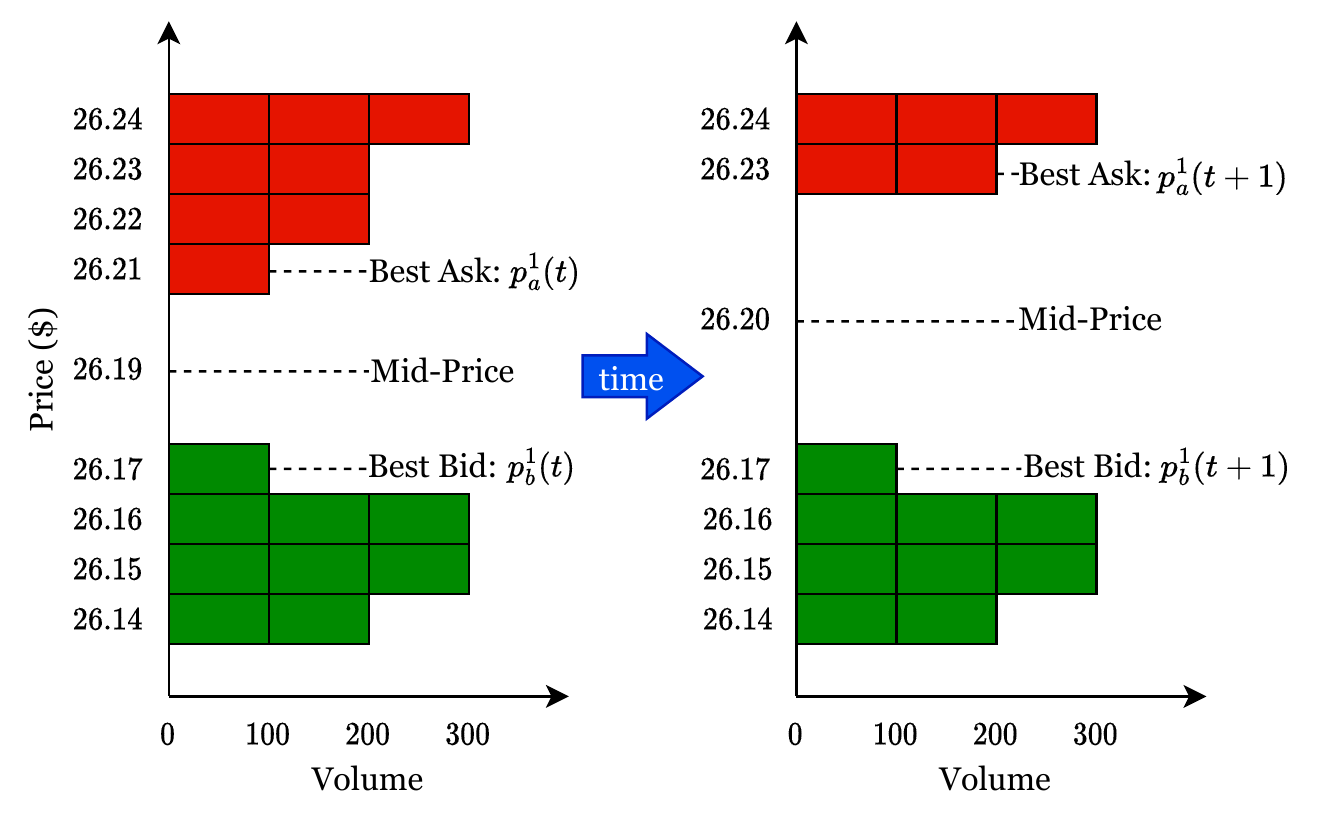}}
\caption{Time evolution of a limit order book from $t$ to $t+1$. After the arrival of a market buy order at $t+1$, the first two levels on the ask side are completely filled and, as a result, the best ask and the mid-price move up.}
\label{fig1}
\end{figure}

A limit order book (LOB) is a collection of all outstanding limit (price-conditional) orders
submitted by market participants. The LOB is arranged into two opposing sides, each with multiple price levels: the \emph{ask-side}, with orders submitted by traders wishing to sell, and the \emph{bid-side}, composed of all buy orders. Fig.~\ref{fig1} shows a simplified visualization of two snapshots of a limit order book. The red (green) rectangles in the diagram represent sell (buy) orders, and their placement within the book depends on the price level specified when submitting a particular order. The lowest price level among all sell orders is known as \emph{best ask} and is denoted by $p_a^1$, while the equivalent on the bid side (\emph{best bid}) corresponds to the highest buy order, denoted by $p_b^1$. The two diagrams of Fig.~\ref{fig1} show what happens when a market buy order (order to buy at the best available price) is submitted at time $t+1$ with volume equal to 300. The bottom two price levels on the ask side are completely filled (since their cumulative volume at time $t$ is equal to the size of the incoming market order) and, as a result, the best ask and the mid-price increase.

Although it is not possible to trade exactly at the mid-price, traders treat it as a proxy for a market value of a stock. There are other important considerations, such as market impact, order sizing and transaction costs, that high-frequency traders need to take into account when submitting their orders. However, the focus of this work is on the central component of this process: correctly predicting the direction of future mid-price that, in turn, allows traders to design profitable strategies based on signals indicating what action should be taken next.

\subsection{Multi-Head Self-Attention}

To set the stage, we first discuss the self-attention mechanism, introduced by \cite{b8} to enable parallelizability in sequential tasks. Axial attention, the key operation within Axial-LOB, is an instance of this concept and is discussed in the next section. The ability of self-attention to directly encode long-range dependencies in data has allowed transformer models to achieve state-of-the-art performance in many areas, including natural language processing, speech and vision. It has also been used in \cite{b9}, in the form of a \emph{multi-head attention} (MHA) layer, for forecasting LOB price changes, as an extension to the methodology developed by \cite{b10}. 

For an input feature map $x\in \mathbb{R}^{C_{in} \times H \times W}$ with number of input channels $C_{in}$, height $H$, and width $W$, the output of a self-attention layer at each position is given by

\begin{equation} \label{eq1}
y_{ij} = \sum_{h=1}^{H} \sum_{w=1}^{W} softmax(q_{ij}^Tk_{hw})v_{hw},
\end{equation}

\noindent where $q=W_Qx$, $k=W_Kx$, and $v=W_Vx$ are the query, key, and value linear projections of the input $x$, computed using learnable matrices $W_Q, W_K, W_V \in \mathbb{R}^{C_{in} \times C_{out}}$. We notice from (\ref{eq1}) that, unlike convolution, the self-attention mechanism utilizes the whole feature map to capture non-local context by pooling values $v$ using global affinities given by the output of the softmax operation. 

We can extend this computation to multi-head attention to capture information from multiple representation subspaces,

\begin{equation} \label{eq2}
MHA = Concatenate(head_1, \cdots, head_H)W^O,
\end{equation}

\noindent where the single-head attention of (\ref{eq1}) is computed $H$ times in parallel using different projection matrices $W_Q^h, W_K^h, W_V^h, \forall h \in \{1, ..., H\}$, and the final output is obtained by concatenating results from each head and linearly projecting with a learnable matrix $W^O$. It has to be noted, however, that self-attention, in its original form, is very expensive to compute, having complexity $\mathcal{O}(h^2w^2)$, which makes it infeasible to apply to high-dimensional inputs.

\subsection{Axial Attention}

To tackle the high computational cost of the na\"ive attention mechanism, one could simply apply local constraints, as proposed by \cite{b11, b12}. However, this work limits the receptive field of the model in a way that is similar to convolutions. As a key part of the Axial-LOB architecture we instead adopt \emph{axial attention} \cite{b13}, which allows us to operate with the global receptive field, while at the same time being computationally efficient. Axial attention factorizes the standard 2D attention into two separate 1D attention modules, a first one that attends to positions along the width axis followed by a second one that pools values along the height axis, as illustrated in Fig.~\ref{fig2}. This sequential operation captures global interactions while reducing the computational complexity to $\mathcal{O}(hwm)$, where $m$ corresponds to all input features.

\subsection{Gated Positional Embeddings}

Axial-LOB incorporates a further extension to the concept of axial attention, that of gated positional embeddings. These were proposed in \cite{b15}, as an extension of the work of  \cite{b14}, which enhances axial attention by incorporating learned relative positional encodings into the attentional affinities. These extra bias terms, in essence, provide the model with a dynamic prior on which parts of the receptive field are most relevant,

\begin{equation} \label{eq3}
y_{ij} = \sum_{h=1}^{H} softmax(q_{ij}^Tk_{hj} + q_{ij}^Tr_{hj}^q + k_{hj}^Tr_{hj}^k)(v_{hj} + r_{hj}^v),
\end{equation}

\noindent where $r^q, r^k, r^v \in \mathbb{R}^{H \times H}$ are the learnable positional encodings for queries, keys, and values, respectively. Since (\ref{eq3}) describes position-sensitive axial attention applied only along the height axis, to obtain the global receptive field, the same operation is also carried out along the width axis.

\begin{figure}[t]
\centerline{\includegraphics[scale=0.5]{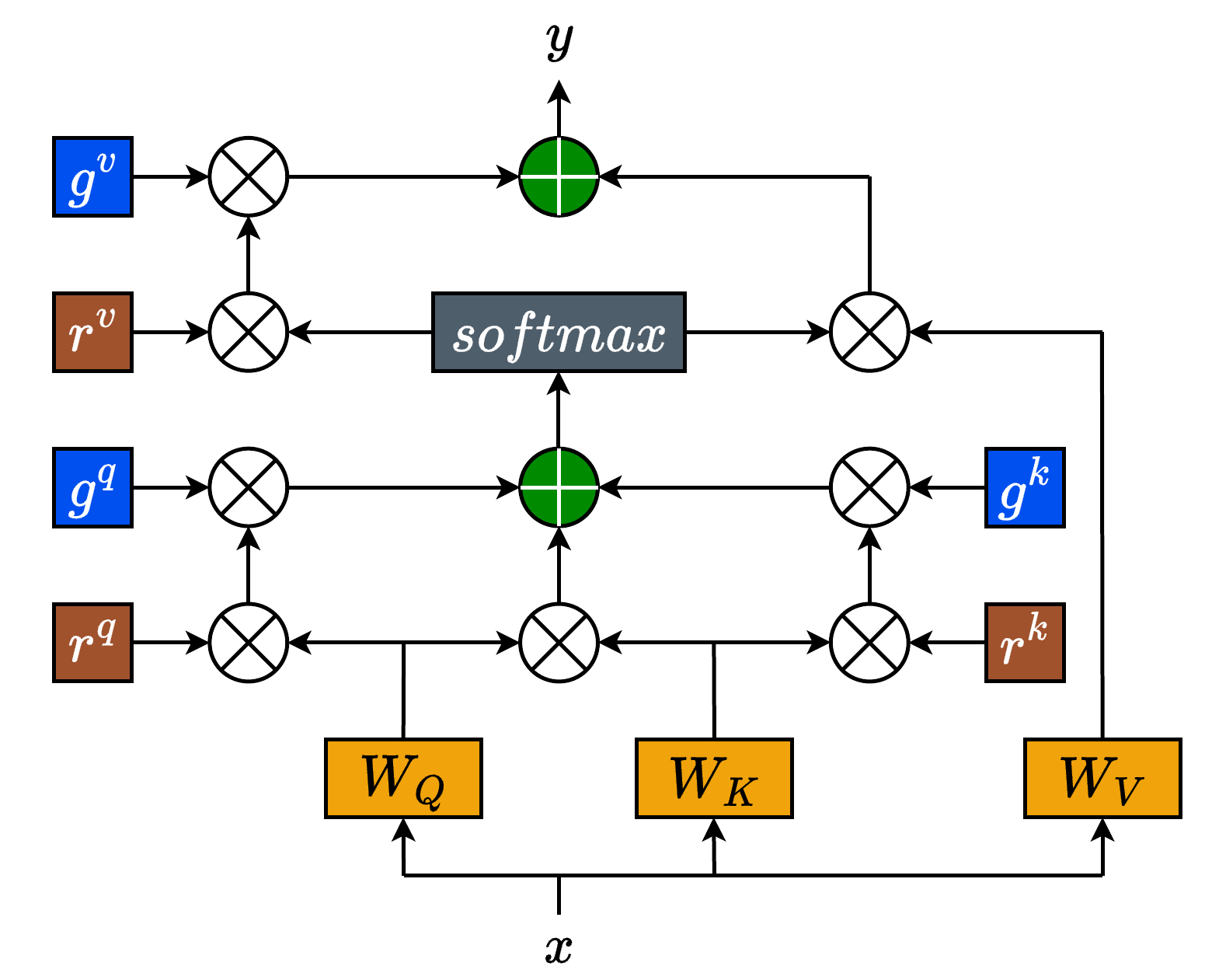}}
\caption{The gated axial attention mechanism sits at the center of the Axial-LOB model and is where the bulk of the computation happens. The influence of the positional encodings (brown) is controlled by the gates (blue).}
\label{fig3}
\end{figure}

As mentioned, we use an extension of (\ref{eq3}), proposed in \cite{b15}, which uses additional gating mechanisms. However, our Axial-LOB model, instead of applying gates to both the value vector and its positional encoding, controls only the information flow from the positional bias terms, as visualised in Fig.~\ref{fig3} and described using the following expression

\begin{equation} \label{eq4}
y_{ij} = \sum_{h=1}^{H} softmax(q_{ij}^Tk_{hj} + g_qb_q + g_kb_k)(v_{hj} + g_vb_v),
\end{equation}

\noindent where $b_q = q_{ij}^Tr_{hj}^q$, $b_k = k_{hj}^Tr_{hj}^k$, $b_v = r_{hj}^v$ and $g_q, g_k, g_v$ are the gating mechanisms, which are learnable parameters. The intuition behind this design choice is analogous to that proposed by \cite{b15}. More specifically, it may be difficult to learn accurate positional bias representations from highly noisy financial data. Therefore, in such circumstances, it becomes advantageous to control, through the use of gates, the level of influence that these positional encodings have on the computation of long-range context.

\section{Methodology}

\subsection{Axial-LOB Architecture} \label{model_architecture}

The architecture of our proposed Axial-LOB model is shown in Fig.~\ref{fig2}. Note that we do not use any technical indicators as part of the feature set; the input contains only historical observations from the limit order book, including price and volume at each level on both bid and ask sides of the book. More specifically, we use the 40 most recent snapshots of the LOB, made up of ten price levels and the corresponding volumes for each side of the book. Therefore, each input to the network can be described as a single-channel image $X \in \mathbb{R}^{H \times W \times 1}$, where the height dimension $H$ represents evolution in time and the width dimension $W$ corresponds to the input features. To be more concrete, $X=[p_a^{(i)}, v_a^{(i)}, p_b^{(i)}, v_b^{(i)}]_{i=1}^{10}$, where $p_a^{(i)}, v_a^{(i)}, p_b^{(i)}, v_b^{(i)} \in \mathbb{R}^{40 \times 1}$ denote the time evolution of ask price, ask volume, bid price, and bid volume, respectively, at the $i$-th LOB level.

The main building component of the proposed model, shown in Fig.~\ref{fig2}, is the gated axial attention block, which consists of two layers, each containing two multi-head axial attention modules with gated positional encodings, the first module acting along the width (feature) dimension, directly followed by the second, which operates along the height (time) dimension. We use $1x1$ convolutions, followed by batch normalization and ReLU activation before and after these consecutive attention operations to adjust the number of channels in the intermediate layers of the network. It should be emphasized that these do not perform any spatial convolutions and are used instead as a way to manage model complexity, by implementing feature map pooling. The number of attention heads in each module and the channel size at each layer are selected during hyperparameter optimization. Additionally, in each layer, the attention maps produced by the top branch are added to the residual connection of the lower branch. Next, adaptive average pooling is applied to feature maps produced by the gated axial attention block and the output is transformed by the fully-connected layer to produce a logit value for each class. Finally, logits are passed through a softmax layer to obtain class probabilities.

\subsection{FI-2010 Benchmark Dataset \& Target Calculation}

The FI-2010 dataset \cite{b16} is a large publicly available benchmark dataset containing high-frequency updates from limit order books of five stocks operating in the Nasdaq Nordic stock market. It is made up of ten consecutive days of trading with price and volume information for the first ten levels on each side of the LOB. FI-2010 is often described as the MNIST or ImageNet of high-frequency trading and previous studies (for example \cite{b4}) have made extensive use of this dataset. It should be noted that we do not utilize any of the hand-crafted features included in the FI-2010 dataset and that we restrict all LOB updates to normal trading hours.

The objective of the Axial-LOB model is to predict future movements of the \emph{mid-price}, as introduced and illustrated by Fig.~\ref{fig1}, and defined using the following form

\begin{equation} \label{eq5}
p(t) = \frac{p_a^1(t) + p_b^1(t)}{2},
\end{equation}

\noindent with the aim of classifying mid-price movement into going up, staying approximately stationary (with respect to a threshold $\alpha$ discussed below), or going down. We note that since events are measured in tick time, not clock time, the interval between consecutive events can vary from a fraction of a second to seconds. In addition, since financial data are highly noisy, a smoothed version of the future mid-price is in practice used, which corresponds to the mean of the next $k$ mid-prices, computed as follows,

\begin{equation} \label{eq6}
m_k^+(t) = \frac{1}{k} \sum_{i=0}^{k} p(t+i),
\end{equation}

\noindent where $k$ is the \emph{prediction horizon}, which can also be thought of as the length of the denoising window, and we conduct experiments using five different values of $k = (10, 20, 30, 50, 100)$. The direction of the price movement is then computed as the proportional change of the smoothed future mid-price with respect to the mid-price observed at $t$:

\begin{equation} \label{eq7}
d_k(t) = \frac{m_k^+(t) - p(t)}{p(t)}.
\end{equation}

\noindent Finally, to obtain the target labels, the price direction $d_k(t)$ is compared with a threshold $\alpha$. We use $\alpha = 0.002$, in order for our results to be comparable to the work of others (benchmark models of Section \ref{section_benchmarks}, which also used this value), and label the price direction as 'up' if $d_k(t) > \alpha$, 'down' if $d_k(t) < -\alpha$ and all other cases as the 'stationary' class. We are aware of other more sophisticated techniques for computing future price direction, but we use the formula in (\ref{eq7}) to establish a fair comparison with the previous studies.

\begin{figure}[t]
\centerline{\includegraphics[scale=0.55]{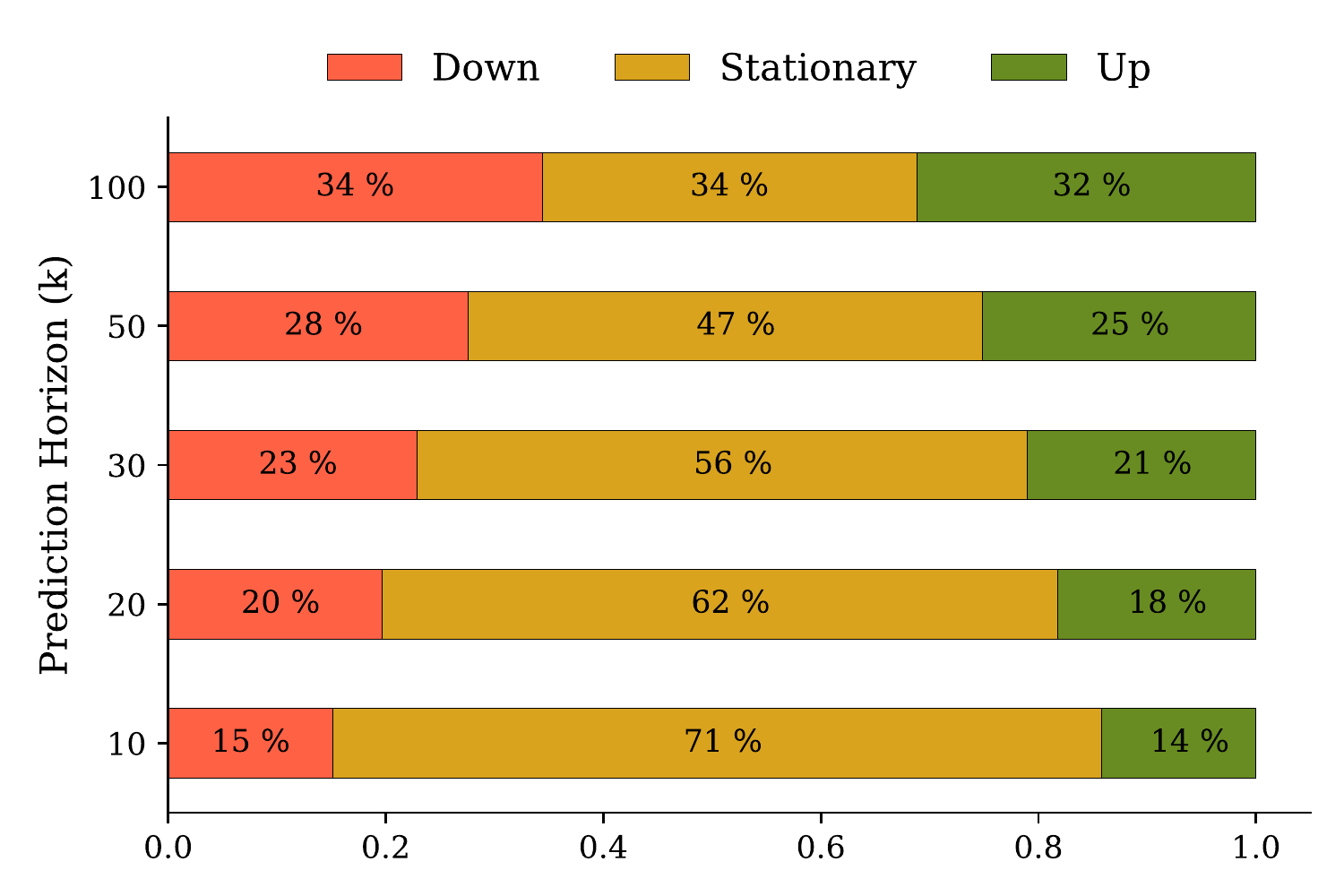}}
\caption{Class distribution at each prediction horizon $k$. As the horizon increases, the labels get more evenly distributed, lowering the class imbalance.}
\label{fig6}
\end{figure}

\subsection{Training \& Model Calibration}

The ten-day FI-2010 dataset is split into training and test segments, where the first seven trading days are used to train the model, and the remaining three days are used as the out-of-sample test data. Additionally, we set aside the last 20\% of the training segment as a separate validation set to tune model hyperparameters using 100 iterations of random grid search. Since the FI-2010 dataset was constructed using high-frequency (millisecond basis) intraday trading events observed exclusively between 10:30 and 18:00 on each specific day, there exists a relatively large gap between the training and the testing segments, preventing any label leakage. For a more detailed description of the dataset construction, the reader is referred to \cite{b16}.

The Axial-LOB network is trained via mini-batch stochastic gradient descent (SGD) by minimising the cross-entropy loss between the predicted class probabilities and the ground truth label,

\begin{equation} \label{eq8}
\mathcal{L}_{CE} = - \sum_{c=1}^C y_{c} \times log \left( \frac{exp(x_{c})}{\sum_{i=1}^C exp(x_{i})} \right),
\end{equation}

\noindent where $x$ is the input to the softmax operator, $y$ is the target label, and the summation is applied over all three classes denoted as $C$. The total loss is then obtained by computing an average of the losses over all training examples. We use momentum \cite{b17}, a batch size of 64, and train the network for 100 epochs. Training of the gating elements is delayed until epoch 5 since this makes their convergence faster and more stable. The learning rate is adjusted at each step using a cosine annealing schedule \cite{b18} that lowers the initial rate as the training progresses using the learning rate multiplier,

\begin{equation} \label{eq9}
LR_{decay} = \frac{1}{2} \times \left( 1 + cos \left( \pi \times \frac{T_{cur}}{T_{total}} \right) \right),
\end{equation}

\noindent where $T_{cur}$ is the current optimizer step and $T_{total}$ corresponds to the total number of training steps over which the cosine decay function is applied. Additionally, to tackle problems related to overfitting, we implement early stopping, where the training of the model is terminated when the validation loss does not improve for 10 consecutive epochs. All experiments are conducted on a single NVIDIA Tesla P100 16GB GPU with 55GB of RAM memory.

\subsection{Benchmark Models} \label{section_benchmarks}

We compare a wide set of benchmarks to our Axial-LOB model: (1) the CNN-based model of \cite{b3}, composed of several convolutional and fully connected layers with temporally aware normalization of the LOB data; (2) an attention-augmented bilinear network (B(TABL)) \cite{b10} with one hidden layer, a bilinear projection layer enhanced by a temporal attention mechanism; (3) an architecture as in (2), but with two hidden layers (C(TABL)) \cite{b10}; (4) a deep neural network with several spatial convolutional layers combined with an Inception Module and followed by LSTM (DeepLOB) \cite{b4}; (5) an encoder-decoder architecture (DeepLOB-Seq2Seq) \cite{b5}, which uses DeepLOB in the encoder block to extract representative features then fed into a simple sequence-to-sequence decoder; and (6) a high-performance extension of the architecture of (5), where the decoder block is implemented using an attention mechanism (DeepLOB-Attention) \cite{b5}, which represents the previous state of the art.

\section{Results}

\subsection{Performance on the FI-2010 Dataset}

\begin{figure*}[t]
\centerline{\includegraphics[scale=0.43]{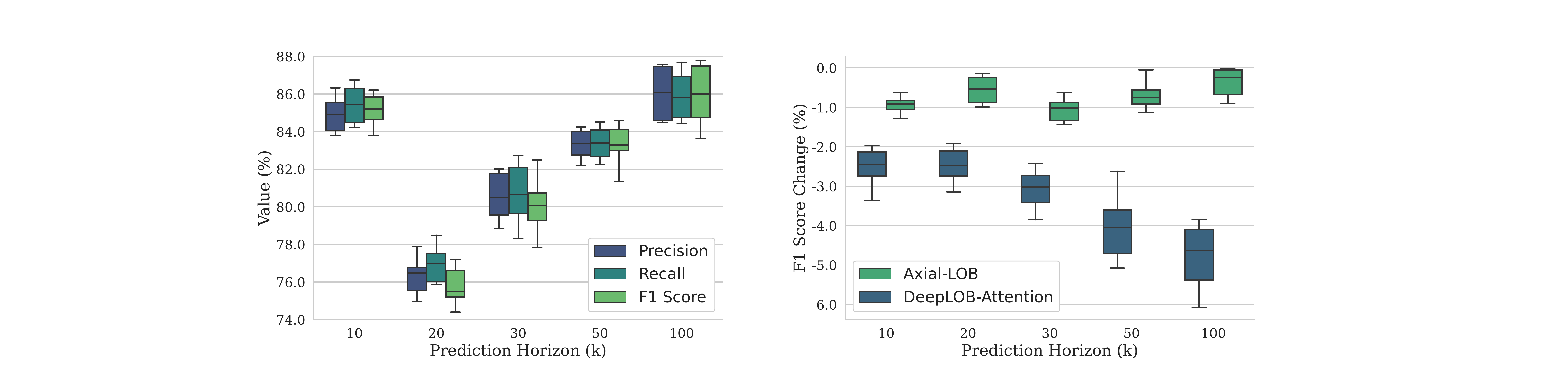}}
\caption{(Left) Boxplot with results from all independent runs of the Axial-LOB model at each prediction horizon $k$. Colors correspond to different evaluation metrics. (Right) Boxplot showing the change in F1 score for our Axial-LOB model (green) and the CNN-based DeepLOB-Attention (blue) at each prediction horizon $k$ under random input feature permutations. Axial-LOB does not rely on a pre-defined order of the input features, delivering a stable performance.}
\label{fig4}
\end{figure*}

The classification performance of Axial-LOB is compared to that of the benchmark algorithms of Section \ref{section_benchmarks} using precision, recall, and F1 score, with the last of these being our main metric as we observe from Fig.~\ref{fig6} that the FI-2010 dataset is imbalanced, especially at short-term prediction horizons, and follow \cite{b16}, who for this reason suggest to focus on F1 score, when conducting model comparisons, on the basis that it combines precision and recall and is robust when dealing with imbalanced class distributions. The experimental results are presented as the left boxplot in Fig.~\ref{fig4}. We evaluate our model using five independent trials due to the stochastic nature of the optimizer and report the means in Table \ref{tab1}.

It can be seen that the Axial-LOB architecture achieves a new state-of-the-art classification performance at all prediction horizons, delivering the highest precision, recall, and F1 score among all benchmark algorithms, beating at each horizon its closest competitor with a statistical significance of $< 5\%$. We note that the performance gap between our model and the other algorithms is relatively small at short prediction horizons ($k=10$ and $k=20$), but becomes wider for bigger values of $k$, with the left plot of Fig.~\ref{fig4} showing a generally upward trend in model performance as the prediction horizon increases, with the shortest horizon ($k=10$) being an exception that could be explained by the assumption that predicting price moves in the near future is an easier task. This trend suggests that Axial-LOB may perform better when there is less imbalance in the class distribution, as observed in Fig.~\ref{fig6}. However, as mentioned previously, the prediction horizon can also be thought of as the length of the denoising window used to calculate the target, so an alternative interpretation of the trend could be that there is less noise in the labels at larger prediction horizons, making the difference between up, down and stationary classes more significant, with the model's performance increasing as a result.

Finally, it is worth pointing out that the Axial-LOB architecture is able to achieve its excellent performance using a relatively low model complexity. Table \ref{tab3} compares the size of the parameter spaces of Axial-LOB and the benchmark models; we note that our network, with a number of parameters of the same order of magnitude as the smallest architectures (B(TABL) \& C(TABL)), is able to outperform the DeepLOB family of models that have the highest complexity.

\begin{table}[t]
\begin{center}
\caption{Performance comparison between our Axial-LOB model and the benchmark algorithms on the FI-2010 dataset.}
\label{tab1}
\begin{tabular}{l|ccc}
\hline
\bfseries Model & \bfseries Precision (\%) & \bfseries Recall (\%) & \bfseries F1 (\%) \\
\hline
\multicolumn{4}{c}{Prediction Horizon $k=10$} \\
\hline
CNN & 50.98 & 65.54 & 55.21 \\
B(TABL) & 68.04 & 71.21 & 69.20 \\
C(TABL) & 76.95 & 78.44 & 77.63 \\
DeepLOB & 84.00 & 84.47 & 83.40 \\
DeepLOB-Seq2Seq & 81.65 & 82.58 & 81.51 \\
DeepLOB-Attention & 82.50 & 83.28 & 82.37 \\
\hline
Axial-LOB & 84.93 & 85.43 & \bfseries 85.14 \\
\hline
\multicolumn{4}{c}{Prediction Horizon $k=20$} \\
\hline
CNN & 54.79 & 67.38 & 59.17 \\
B(TABL) & 63.14 & 62.25 & 62.22 \\
C(TABL) & 67.18 & 66.94 & 66.93 \\
DeepLOB & 74.06 & 74.85 & 72.82 \\
DeepLOB-Seq2Seq & 73.12 & 74.38 & 72.99 \\
DeepLOB-Attention & 74.31 & 75.25 & 73.73 \\
\hline
Axial-LOB & 76.32 & 76.98 & \bfseries 75.78 \\
\hline
\multicolumn{4}{c}{Prediction Horizon $k=30$} \\
\hline
CNN & 66.52 & 67.98 & 65.72 \\
B(TABL) & 70.13 & 69.76 & 67.08 \\
C(TABL) & 72.90 & 71.88 & 69.34 \\
DeepLOB & 76.00 & 76.36 & 75.33 \\
DeepLOB-Seq2Seq & 75.86 & 76.41 & 75.75 \\
DeepLOB-Attention & 77.32 & 77.59 & 76.94 \\
\hline
Axial-LOB & 80.54 & 80.69 & \bfseries 80.08 \\
\hline
\multicolumn{4}{c}{Prediction Horizon $k=50$} \\
\hline
CNN & 55.58 & 67.12 & 59.44 \\
B(TABL) & 74.58 & 73.09 & 73.64 \\
C(TABL) & 79.05 & 77.04 & 78.44 \\
DeepLOB & 80.38 & 80.51 & 80.35 \\
DeepLOB-Seq2Seq & 77.96 & 78.10 & 77.99 \\
DeepLOB-Attention & 79.51 & 79.49 & 79.38 \\
\hline
Axial-LOB & 83.31 & 83.38 & \bfseries 83.27 \\
\hline
\multicolumn{4}{c}{Prediction Horizon $k=100$} \\
\hline
CNN & 65.51 & 64.87 & 65.05 \\
B(TABL) & 70.08 & 68.59 & 69.14 \\
C(TABL) & 75.55 & 73.54 & 74.94 \\
DeepLOB & 76.85 & 76.72 & 76.76 \\
DeepLOB-Seq2Seq & 79.31 & 79.09 & 79.16 \\
DeepLOB-Attention & 81.62 & 81.45 & 81.49 \\
\hline
Axial-LOB & 86.04 & 85.92 & \bfseries 85.93 \\
\hline
\end{tabular}
\end{center}
\end{table}

\begin{table}[htbp]
\begin{center}
\caption{Comparison of model complexities (expressed as the number of parameters) between Axial-LOB and the benchmarks.}
\label{tab3}
\begin{tabular}{l|c}
\hline
\bfseries Model & \bfseries \# of model parameters \\
\hline
CNN & 17,635 \\
B(TABL) & 5,844 \\
C(TABL) & 11,344 \\
DeepLOB & 142,435 \\
DeepLOB-Seq2Seq & 176,419 \\
DeepLOB-Attention & 177,699 \\
\hline
Axial-LOB & 9,615 \\
\hline
\end{tabular}
\end{center}
\end{table}

\subsection{Performance under Input Permutation}

Results of Table \ref{tab1} show that Axial-LOB outperforms all the CNN-based architectures, including the previous state of the art, DeepLOB-Attention, which applies the attention mechanism to features extracted from data using carefully hand-crafted spatial convolutional filters. DeepLOB-Attention network operates in sequential steps, first summarizing information from price-volume pairs at each LOB level and then aggregating information across multiple levels. In order to effectively scan through the input data, this approach requires that the shapes of convolutional filters and strides match a pre-defined input layout. In contrast, Axial-LOB builds attentional affinities directly from raw LOB data, which allows the model to remain agnostic with respect to the order of the input features. This proposition is tested by selecting one set of starting weights but carrying out five independent experimental trials, each using a different random permutation of the input features, with a single input at time $t$ defined as $x_t=[f^{(1)}_t, f^{(2)}_t, \dots, f^{(i)}_t, \dots, f^{(40)}_t]$, where $f^i$ denotes price or volume at any of the first ten limit order book levels. Note that this is in contrast to the data structure of Section \ref{model_architecture}, where the input was composed of price-volume pairs in ascending LOB-level order.

Table \ref{tab2} reports the means and standard deviations of the changes in F1 score, for Axial-LOB and the previous state of the art, DeepLOB-Attention. Results are also presented as a boxplot in the right section of Fig.~\ref{fig4}. The CNN-based DeepLOB-Attention model displays a relatively large drop in F1 score, more noticeable as the prediction horizon increases, while the performance of our Axial-LOB model remains relatively stable. This is because, unlike the CNN-based architectures, Axial-LOB does not rely on a pre-defined order of features that has to match the shape of the convolutional kernels and strides employed by those models. Therefore, results suggest that our model can work equally well with different input permutations and so has the capacity to more easily incorporate additional LOB features without the need to redesign the network architecture.

\begin{table}[htbp]
\begin{center}
\caption{Means and standard deviations of the changes to F1 score under input permutation.}
\label{tab2}
\begin{tabular}{c|c|c}
\hline
\bfseries Prediction Horizon & \bfseries Axial-LOB & \bfseries DeepLOB-Attention \\
\hline
$k=10$ & $-0.94 \pm 0.22$ & $-2.53 \pm 0.49$ \\
$k=20$ & $-0.56 \pm 0.33$ & $-2.48 \pm 0.44$ \\
$k=30$ & $-1.05 \pm 0.30$ & $-3.09 \pm 0.50$ \\
$k=50$ & $-0.68 \pm 0.36$ & $-4.01 \pm 0.86$ \\
$k=100$ & $-0.37 \pm 0.35$ & $-4.81 \pm 0.83$ \\
\hline
\end{tabular}
\end{center}
\end{table}

\section{Conclusions}

This work has introduced the Axial-LOB model, used to predict mid-price movements of stocks using limit order book data. The proposed deep-learning architecture is based on a gated position-sensitive axial attention mechanism that allows our model to efficiently operate with the global receptive field. To our knowledge, this is the first application of axial attention to financial data. Results demonstrate that Axial-LOB achieves excellent classification performance on the FI-2010 dataset, outperforming all benchmark algorithms and establishing a new state of the art, while in addition using a much lower model complexity. Unlike previous CNN-based deep learning approaches to stock prediction from LOB data, our architecture does not require any hand-crafted spatial convolutional kernels, thus delivering stable performance under different input feature permutations.

In future work, we plan to further test the robustness and generalization ability of the Axial-LOB model using a larger high-frequency dataset. Another interesting research direction would be to extend the current model architecture by adding a decoder block that would simultaneously generate predictions for all horizons.

\end{document}